\documentclass[reprint,superscriptaddress,showpacs,nofootinbib,amsmath,amssymb,aps,prl]{revtex4-1}
\usepackage{graphicx}
\usepackage{dcolumn}
\usepackage{bm}
\usepackage[utf8]{inputenc}
\usepackage[frenchb]{babel}
\usepackage{braket}

\begin{document}

\preprint{APS/123-QED}

\title{Dipolar polaritons squeezed at unitarity}

\author{S. V. Andreev}
\email[Electronic adress: ]{Serguey.Andreev@gmail.com}
\affiliation{National Research Center "Kurchatov Institute" B.P.\ Konstantinov Petersburg Nuclear Physics Institute, Gatchina 188300, Russia}

\date{\today}

\begin{abstract}  
Interaction of dipolar polaritons can be efficiently tuned by means of a shape resonance in their excitonic component. Provided the resonance width is large, a squeezed population of strongly interacting polaritons may persist on the repulsive side of the resonance. We derive an analytical expression for the polariton coupling constant and show that it may have very large values in typical experimental conditions. Our arguments provide a new direction for the quest of interactions in quantum photonics.   
      
\end{abstract}


\maketitle

The commonly adopted strategy to introduce interactions into the quantum optics of semiconductors is tailoring the non-linearity due to excitonic transitions \cite{Ciuti}. In the regime of strong light-matter coupling the macroscopic population of the cavity mode is efficiently transferred into the exciton field, which can be regarded as a gas of bosonic quasiparticles \cite{Keldysh, Hanamura, Littlewood, Amand}. In particular, the blueshift of the polariton dispersion is governed by low-energy s-wave collisions between the pairs of excitons \cite{Keldysh}. This naturally refers to ultra-cold atomic systems, where enhancement of interactions has been demonstrated by working with species having dipole moments \cite{Lahaye}, Rydberg excitations \cite{Low} and using the technique of Feshbach resonance \cite{Feshbach}. The latter provides a possibility to tune the scattering length from positive to negative values through the unitary limit by adjusting the position of the scattering threshold with respect to a bound state.

On the technological side, exceptional excitonic properties are found in atomically thin heterostructures of transition metal dichalcogenides (TMD's) \cite{Colloquium}. The atom-like Lennard-Jones interaction between excitons has been demonstrated in these materials \cite{LennardJones}. Such interaction naturally admits a bound state and, indeed, biexcitons have recently been observed in several types of monolayers \cite{XXinTMD}. A fundamental difference from the atomic clouds is, however, a purely two-dimensional (2D) character of the exciton translational motion. The interactions in a 2D ultra-cold gas are generically weak due to the properties of 2D kinematics. Thus, in contrast to three dimensions, quantum scattering off a weakly-bound state has a vanishingly small amplitude \cite{Landau}. At sufficiently low exciton densities these arguments apply also for semiconductor quantum wells (QW's). 

As was proposed by the author \cite{RP}, a 2D analog of the Feshbach resonance may be realized with dipolar excitons formed of electrons and holes residing in spatially separated layers. The dipolar repulsion introduces a potential barrier between the outer continuum and the bound state (biexciton), which enables a quasi-discrete level with tunable energy and lifetime. Both parameters can be controlled by changing the distance $d$ between the layers. The attractive side of such resonance was theoretically explored in the context of roton-maxon excitations and supersolidity in dipolar Bose-Einstein condensates (BEC's) \cite{Andreev2015, Andreev2017}. On the repulsive side, the equilibrium ground state is a condensate of biexcitons, distinguished from the exciton condensate by suppressed coherence of the photoluminescence (PL) and a gapped excitation spectrum \cite{RP}. These predictions hold for a wide variety of bilayer structures, where the exciton lifetime is sufficiently long to establish a thermodynamic equilibrium. Thus, the numerical calculations of the exciton interaction potential in coupled QW's \cite{Zimmerman} suggest that the shape resonance may be responsible for the formation of a fragmented-condensate solid of excitons \cite{Andreev2013, Andreev2014, Andreev2015, Andreev2017}.

Several groups have recently reported an increase of the polariton interaction due to the dipolar moment in the excitonic component \cite{Savvidis, Rapaport, Imamoglu}. The results presented in Ref. \cite{Rapaport} are particularly compelling: a factor of 200 enhancement of the dipolar polariton interaction strength as compared to unpolarized polaritons has been detected. Dipolar repulsion alone cannot explain such tremendous blueshift of the polariton PL. The mystery is deepened by very low values of polariton densities at which the experiment was done.

Motivated by these experimental observations, the paper presents a phenomenological model of resonantly paired dipolar polaritons. In contrast to dipolar excitons, microcavity polaritons are far from the thermodynamic equilibrium, their statistics being closer to lasers rather than to atomic BEC's \cite{PolaritonStatistics}. This makes possible existence of a metastable polariton population on the repulsive side of the shape resonance. Coupling to a transient bipolariton mode in this case yields divergent behaviour of the 2D effective interaction, akin to the unitary limit in three-dimensional atomic clouds. We derive an analytical expression for the interaction enhancement factor as a function of the polariton dipole moment and density, and show that it can be very large in typical experimental conditions. Being exact in the dilute regime, this result ultimately holds for two polaritons in vacuum. Another interesting prediction of our theory is that the many-body polariton states become \textit{squeezed} by the resonance. This could be verified in current experiments by examining  the statistics of emitted photons.

Let us discuss the relevant timescales of the problem. First, we shall assume that the polaritons do not condense into the paired state, which is the equilibrium ground state when the discrete level $\varepsilon$ is below the scattering threshold. Second, the width of the resonance $\beta$ must be sufficiently large for polaritons could feel the interior of the barrier during their lifetime $\tau$. We shall assume a \textit{broad} resonance, which seems to be the most likely for polaritons because of their very low effective mass \cite{SI}. Hence, we let
\begin{equation}
\label{timescales}
\beta\gg \varepsilon
\end{equation}
and $\hbar/\beta\ll\tau \ll \tau_{\bm p}$, where $\tau_{\bm p}$ is the thermalization time \cite{footnote}.  

The system is a mixture of two polariton flavours $\hat c_\sigma$ with $\sigma=(\uparrow,\downarrow)$ and their bipolaritonic pairs $\hat C$. In practice, "$\uparrow$" and "$\downarrow$" typically correspond to left- and right-circularly polarized photons \cite{Schneider}. In planar dielectric microcavities the lower polariton dispersion is split into the linearly polarized TM and TE modes \cite{Kaliteevskii}. This splitting acts as an effective magnetic field which flips the polariton pseudospin $\sigma$. Provided that
\begin{equation}
\label{SOC}
p\ll \sqrt{m_\mathrm{LT}\beta}/\hbar,
\end{equation}
the decay of a bound state due to the polariton spin-flip is subdominant with respect to the tunneling under the dipolar potential barrier. Here, the parameter $m_\mathrm{LT}$ accounts both for the bare photonic and low-$k$ excitonic TE-TM splitting \cite{Sham, Glazov}. In confined geometries \cite{Lundt, Bloch} one should let $p\sim \pi/L$, where $L$ is the characteristic length of the confinement. Under the condition \eqref{SOC}, we may consider a single polariton branch $E(\bm p)$ characterized by the effective mass $m$ in the vicinity of its minimum. Analysis of a possible departure from this approximation will be given in a separate paper. 

The many-body Hamiltonian reads
\begin{widetext}
\begin{equation}
\label{Hamiltonian}
\begin{split}
\hat H=&\sum_{\textbf{p},\sigma}E(\bm p)\hat c_{\sigma, \textbf{p}}^{\dag} \hat c_{\sigma, \textbf{p}}+\sum_\textbf{k}[2E(\bm k/2)+\varepsilon]\hat C_\textbf{k}^{\dag} \hat C_\textbf{k}+\frac{g}{2S}\sum_{\textbf p_1,\textbf p_2,\textbf{q},\sigma}\hat c_{\sigma, \textbf p_1+\textbf q}^{\dag} \hat c_{\sigma,\textbf p_2-\textbf q}^{\dag}\hat c_{\sigma, \textbf p_1}\hat c_{\sigma,\textbf p_2}\\
&+\sqrt{\frac{\hbar^2\beta}{2\pi m S}}\sum_{\bm k, \bm p}\left(\hat c_{\uparrow,\bm p+\frac{\bm k}{2}}^\dagger \hat c_{\downarrow,-\bm p+\frac{\bm k}{2}}^\dagger \hat C_{\bm k}+ \hat c_{\uparrow,-\bm p+\frac{\bm k}{2}} \hat c_{\downarrow,\bm p+\frac{\bm k}{2}} \hat C_{\bm k}^\dagger\right).
\end{split}
\end{equation}
\end{widetext}
Here the first two terms describe the dispersions of single- and bipolaritons, respectively, with $\bm p=(p_x,p_y)$ and $E(\bm p)=\hbar^2 p^2/2m$ at the bottom of the band ($\bm p\rightarrow 0$). In the limit of zero exciton dipole moment the discrete level $\varepsilon$ can be identified with the binding energy of a loosely bound bipolariton molecule \cite{Rocca}. The next two terms is the usual background interaction between the polaritons with alike spins (accounting both for the short-range part and the dipolar tail of the bare exciton potential) in the quantization area $S$. The last term models the interaction of polaritons with opposite spins by converting them into the bipolariton mode and vice versa. The square-root prefactor is constructed in such a way as to reproduce the low-energy 2D scattering amplitude for two particles in vacuum \cite{RP, PolaritonicFR}.

By using the standard commutation relations for bosons, one obtains the following set of Heisenberg equations of motion
\begin{widetext}
\begin{subequations}\label{Heisenberg}
\begin{align}
\label{Heisenberg1}
i\hbar \frac{d\hat c_{\sigma, \bm p}}{dt}&=[E(\bm p)+\mu_\sigma]\hat c_{\sigma, \bm p}+\sqrt{\frac{\hbar^2\beta}{2\pi m S}}\sum_{ \bm k}\hat c_{\sigma^\prime\neq\sigma, \bm{k}}^{\dag}\hat C_{\bm k+\bm p}\\
\label{Heisenberg2}
i\hbar \frac{d\hat C_{\bm k}}{dt}&=[2E(\bm k/2)+\varepsilon]\hat C_{\bm k}+\sqrt{\frac{\hbar^2\beta}{2\pi m S}}\sum_{\bm p} \hat c_{\uparrow,-\bm p+\frac{\bm k}{2}} \hat c_{\downarrow,\bm p+\frac{\bm k}{2}},
\end{align}
\end{subequations}
\end{widetext}
where we have replaced the Hartree groups of operators by $c$-numbers and defined
\begin{equation}
\label{mu}
\mu_\sigma=\frac{g}{S}\sum_{\bm q} \lvert c_{\sigma, \textbf{q}} \rvert^2=gn_\sigma,
\end{equation}
with $n_\sigma$ being the polariton densities in each component. By introducing the slowly-varying amplitudes
\begin{equation}
\label{amplitudes}
\hat c_{\sigma, \bm p}=\mathsf{\hat c}_{\sigma, \bm p}e^{-i[E(\bm p)+\mu_\sigma] t/\hbar},
\end{equation}
we notice existence of a stationary ($ d\mathsf{\hat c}_{\sigma, \bm p}/dt=0$) solution of Eq. \eqref{Heisenberg2} in the form
\begin{equation}
\label{bipolariton}
\hat C_{\bm k}=\frac{\sqrt{\frac{\hbar^2\beta}{2\pi m S}}}{\mu_\uparrow+\mu_\downarrow-\varepsilon^\prime_{\bm k}}\sum_{\bm p} \hat c_{\uparrow,-\bm p+\frac{\bm k}{2}} \hat c_{\downarrow,\bm p+\frac{\bm k}{2}},
\end{equation}
where the motion of the bipolariton mode is reduced to that of a pair of polaritons with opposite spins. Here $\varepsilon^\prime_{\bm k}=\varepsilon-\varepsilon_{\bm k}$ with
\begin{equation}
\label{Erel} 
\varepsilon_{\bm k}=\frac{\sum_{\bm p} E_\mathrm{rel} (\bm p, \bm k) <\mathsf{\hat c}_{\uparrow,-\bm p+\frac{\bm k}{2}} \mathsf{\hat c}_{\downarrow,\bm p+\frac{\bm k}{2}}>}{\sum_{\bm p}<\mathsf{\hat c}_{\uparrow,-\bm p+\frac{\bm k}{2}} \mathsf{\hat c}_{\downarrow,\bm p+\frac{\bm k}{2}}>}
\end{equation}
being the kinetic energy of the relative motion in the pair, $E_\mathrm{rel}(\bm p, \bm k)=E(\bm p+\bm k/2)+E(-\bm p+\bm k/2)-2E(\bm k/2)$.

The condition \eqref{timescales} provides a physical meaning to the solution \eqref{bipolariton}. The objects $\hat C_{\bm k}$ should be regarded as auxiliary fields describing onset of pair correlations between the polaritons, rather than new (quasi)particles. Indeed, substituting \eqref{bipolariton} into the last term of the Hamiltonian \eqref{Hamiltonian}, one obtains an effective model
\begin{equation}
\label{HamiltonianPrime}
\begin{split}
\hat H^\prime=&\sum_{\textbf{p},\sigma}E(\bm p)\hat c_{\sigma, \textbf{p}}^{\dag} \hat c_{\sigma, \textbf{p}}+\\
&\frac{1}{2S}\sum_{\textbf p_1,\textbf p_2,\textbf{q},\sigma,\sigma^\prime}\hat c_{\sigma, \textbf p_1+\textbf q}^{\dag} \hat c_{\sigma^\prime,\textbf p_2-\textbf q}^{\dag}g_{\sigma \sigma^\prime}\hat c_{\sigma, \textbf p_1}\hat c_{\sigma^\prime,\textbf p_2},
\end{split}
\end{equation}
with $g_{\uparrow\uparrow}=g_{\downarrow\downarrow}=g$ and
\begin{equation}
\label{interspecies}
g_{\uparrow\downarrow}=\frac{\hbar^2}{2\pi m}\frac{\beta}{(\mu_\uparrow+\mu_\downarrow-\varepsilon^\prime_{\bm p_1+\bm p_2})}.
\end{equation}
One can see, that the pair correlations between the polaritons manifest themselves as resonantly strong two-body interactions. For a broad resonance considered in this work, large magnitude of $g_{\uparrow\downarrow}$ is primarily due to the relation \eqref{timescales} (the $\mu_\sigma$'s are on the order of $\varepsilon$). However, by reducing the difference $\mu_\uparrow+\mu_\downarrow-\varepsilon^\prime_{\bm p_1+\bm p_2}$ (by, e.g., decreasing the polariton density) one may also observe the typical resonant growth of the interaction strength. Though here we have in mind the case $\mu_\uparrow+\mu_\downarrow>\varepsilon^\prime_{\bm p_1+\bm p_2}$, the formula \eqref{interspecies} can be used on the attractive side $\mu_\uparrow+\mu_\downarrow<\varepsilon^\prime_{\bm p_1+\bm p_2}$ as well.
 
\begin{figure}[t]
\label{Fig1}
\includegraphics[width=1\columnwidth]{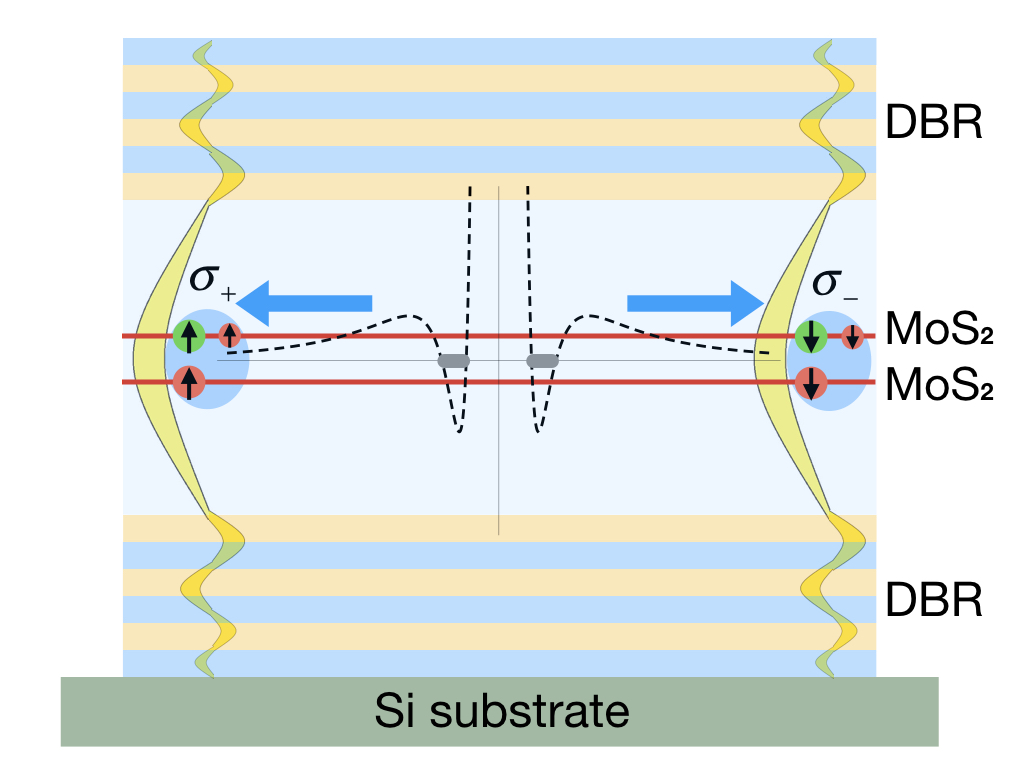}  
\caption{Sketch of a TMD-based planar microcavity featuring unitary dipolar polaritons. The active region consists of a homobilayer structure with AA$^\prime$ or AB stacking, which allows dipolar excitons with strong oscillator strengths and convenient selection rules \cite{Urbaszek, Koch}. Black arrows show the spin orientation of the conduction and valence electronic states in the $K^\pm$ valleys. The "$\uparrow$" and "$\downarrow$" excitons couple to the cavity modes with opposite helicities. Dashed line traces the profile of the two-body interaction potential $V_{\uparrow\downarrow}(r)$ felt by each exciton \cite{SI}. This supports a quasi-discrete level (gray bar) separated from the outer continuum by a barrier due to the dipolar repulsion. The resulting polaritons experience strong repulsion and squeezing of their collective states.}
\end{figure}                

Assume now a resonant excitation of a coherent mixture of "$\uparrow$" and "$\downarrow$" polaritons at some point $\bm p_0$ on the dispersion curve $E(\bm p)$. The output signal is registered at the distance $l\sim \upsilon_{gr} \tau$ from the excitation spot. Here  $\upsilon_{gr}$ is the corresponding group velocity. For simplicity assume equal densities $n_\uparrow=n_\downarrow\equiv n=N/S$, which in practice may be achieved by using linearly polarized light. We therefore let the following initial ($t=0$) configuration:
\begin{equation}
\label{polaritons0}
\begin{split}
&\hat c_{\sigma, p_0}(0)\ket{\psi}=\sqrt{N}\ket{\psi}\\
&\hat c_{\sigma, p\neq p_0}(0)\ket{\psi}=0.
\end{split}
\end{equation}
The bipolariton part of the many-body wavefunction $\ket{\psi}$ is initially in the vacuum state:  
\begin{equation}
\label{bipolariton0}
\hat C_{2p_0}(0)\ket{\psi}=0.
\end{equation}
By substituting the slowly varying $c$-numbers [see the definition \eqref{amplitudes}] $\mathsf c_{\sigma, p_0}=\rho_\sigma e^{i\phi_\sigma}$ and $\mathsf C_{2p_0}=\rho e^{i\phi}$ into Eqs. \eqref{Heisenberg}, and omitting the terms scaling as $\sqrt{\varepsilon/\beta}$, one can find
\begin{equation}
\label{phases}
\begin{split}
 \phi(t)&=\phi_\uparrow(t)+\phi_\downarrow(t)\pm \pi/2\\
 \phi_\sigma(t)&=\phi_\sigma(0)
 \end{split}
\end{equation}
and
\begin{equation}
\begin{split}
\label{magnitudes}
\rho_\sigma (t)&=\sqrt{N} \cosh^{-1}(t/\tau_0)\\
\rho(t)&=\sqrt{N}\tanh(t/\tau_0).
\end{split}
\end{equation}
Eq. \eqref{magnitudes} shows that on the characteristic time scale
\begin{equation}
\label{tau0}
\tau_0=\sqrt{\frac{2\pi m}{\beta n}}
\end{equation}
the coherent mixture is entirely converted into the paired state \eqref{bipolariton}. According to \eqref{HamiltonianPrime} and \eqref{interspecies}, the modified polariton blueshift is given by
\begin{equation}
\label{blueshift}
\mu_\sigma^\prime=gn+\frac{\hbar^2n}{2\pi m}\frac{\beta}{(2ng-\varepsilon)}.
\end{equation}
Note, that spreading of the polariton distribution over the $\bm k$-space region where the dispersion can be approximated by a linear function will not affect the result \eqref{blueshift} since, according to \eqref{Erel}, in this region one has $\varepsilon_{\bm k}\equiv 0$.

The relation \eqref{timescales} refers to the typical case $d\sim d_c$, where $d_c$ is the critical value at which the true bound state disappears \cite{dc}. In general, the width of the resonance changes from $0$ to $\infty$ (the latter describing the ultimate case where the level washes out) as the exciton dipole moment is tuned from $d\ll d_c$ to $d\gg d_c$. For $0\leqslant d\lesssim d_c$ one may take $\beta(d)=\mathcal{B} d$. The corresponding dependence for the position of the level has the form \cite{RP} $\varepsilon(d)=\mathcal{E}(d-d_c)$.

Following Ref. \cite{Rapaport}, one may then consider the "interaction enhancement factor" $\eta(d,n)=\mu_\sigma^\prime/\mu_\sigma-1$
as a function of the dipole moment $d$ and density $n$. Substituting the above relations for $\beta(d)$ and $\varepsilon(d)$ into Eq. \eqref{blueshift}, we obtain
\begin{equation}
\label{eta} 
\eta(d,n)=\frac{\hbar^2}{2\pi m g}\frac{\mathcal{B} d}{[2ng-\mathcal{E}(d-d_c)]}.
\end{equation}
Though our analytical methods do not allow us to estimate the typical values of the parameters $\mathcal{B}$ and $\mathcal{E}$, by virtue of \eqref{timescales} one may expect $\mathcal{B}\gg \mathcal{E}$ and, consequently, $\eta\gg 1$. The relation \eqref{timescales}, in turn, is guaranteed by very low values of the polariton mass $m$ as compared to bare excitons \cite{SI}.  

Interestingly, the strong correlations in the paired state \eqref{bipolariton} squeeze the polariton wavefunctions. To illustrate this point, consider again the situation discussed above, where one starts from a coherent state \eqref{polaritons0} for polaritons and a vacuum state \eqref{bipolariton0} for their pairs. Introduce rotated quadratures
\begin{equation}
\label{polaritonQ}
\begin{split}
\hat x_\sigma&=\tfrac{1}{2}(\mathsf{\hat c}_{\sigma,p_0}e^{-i\phi_\sigma}+\mathsf{\hat c}_{\sigma,p_0}^\dagger e^{i\phi_\sigma})\\
\hat y_\sigma&=\tfrac{1}{2i}(\mathsf{\hat c}_{\sigma,p_0}e^{-i\phi_\sigma}-\mathsf{\hat c}_{\sigma,p_0}^\dagger e^{i\phi_\sigma})
\end{split}
\end{equation}
and 
\begin{equation}
\begin{split}
\hat X&=\tfrac{1}{2}(\mathsf{\hat C}_{2p_0}e^{-i\phi}+\mathsf{\hat C}_{2p_0}^\dagger e^{i\phi})\\
\hat Y&=\tfrac{1}{2i}(\mathsf{\hat C}_{2p_0}e^{-i\phi}-\mathsf{\hat C}_{2p_0}^\dagger e^{i\phi}).
\end{split}
\end{equation}
Write $\hat x_\sigma=x_\sigma+\delta \hat x_\sigma$ and the same for $\hat y_\sigma$, $\hat X$, $\hat Y$. The linearized equations of motion for the quadrature fluctuations read
\begin{equation}
\label{quadraturesT}
\begin{split}
\frac{d}{dt}\delta \hat x_{\uparrow,\downarrow}&=\pm\sqrt{\frac{\beta}{2\pi m S}}(\rho_{\downarrow,\uparrow}\delta \hat X+\rho \delta\hat x_{\downarrow,\uparrow})\\          
\frac{d}{dt}\delta\hat y_{\uparrow,\downarrow}&=\pm\sqrt{\frac{\beta}{2\pi m S}}(\rho_{\downarrow,\uparrow}\delta \hat Y-\rho \delta\hat y_{\downarrow,\uparrow})\\
\frac{d}{dt}\delta \hat X&=\mp\sqrt{\frac{\beta}{2\pi m S}}(\rho_{\uparrow}\delta \hat x_\downarrow+\rho_\downarrow \delta\hat x_{\uparrow})\\
\frac{d}{dt}\delta \hat Y&=\mp\sqrt{\frac{\beta}{2\pi m S}}(\rho_{\uparrow}\delta \hat y_\downarrow+\rho_\downarrow \delta\hat y_{\uparrow}),
\end{split}
\end{equation}
where the sign "$+$" or "$-$" corresponds to the two possible choices of the phase shift in Eq. \eqref{phases}, and $\rho$, $\rho_\sigma$ are given by \eqref{magnitudes}. At $t=0$ one can use Eqs. \eqref{polaritonQ} and \eqref{polaritons0} to find $\braket{\delta \hat x_\sigma^2(0)}=1/4$ and $\braket{\delta \hat y_\sigma^2(0)}=1/4$, the well-known property of a coherent state \cite{Scully}. In contrast, at $\tau_0\ll t \lesssim \tau$, where $\tau_0$ is given by Eq. \eqref{tau0}, one can substitute $\rho_\sigma=0$ and $\rho=\sqrt{N}$ into the first pair of Eqs. \eqref{quadraturesT} to obtain
\begin{equation}
\label{squeezing}
\begin{split}
\braket{\delta \hat x_\sigma^2(t)}&\sim e^{\pm t/\tau_0}\\
\braket{\delta \hat y_\sigma^2(t)}&\sim e^{\mp t/\tau_0},
\end{split}
\end{equation}
showing that the polaritons exhibit 100 $\%$ squeezing in either of the two quadratures at the output.

The requirement $\tau_0\ll \tau$ sets the lower value of the polariton density at which the predicted effects may be observed. For sufficiently large $\beta$ one may operate in the ultra-dilute regime and even in the few-particle limit. In Eqs. \eqref{interspecies},\eqref{blueshift} and \eqref{eta} the latter is formally achieved by letting $n=0$. Strong repulsive interactions between just two polaritons may be particularly promising for implementation of the dual-rail polariton logic \cite{Lisyansky}.

The condition \eqref{timescales} has allowed us to neglect the dissipation and make our arguments particularly transparent. Pair-breaking events due to leakage of the single photons from the cavity result in loss of correlations and, at a first glance, would reduce the degree of squeezing. In practice, however, this reduction may be fully compensated by the noise of the external vacuum (see Ref. \cite{Yurke}), which restores the significance of the result \eqref{squeezing}.

Our last remark concerns the choice of the sign in Eq. \eqref{phases}. Under the condition \eqref{timescales}, the Josephson coupling of the polariton states to the resonance stabilizes a definite phase relation during the signal propagation. The initial configuration is, however, chosen stochastically and may vary from one laser pulse to another. This circumstance should be taken into account when verifying the prediction \eqref{squeezing} experimentally.

Besides the already mentioned state-of-the-art in QW microcavities \cite{Savvidis, Rapaport, Imamoglu}, in Fig. (1) we sketch a possible implementation of unitary polaritons with TMD's. A convenient choice may be a homobilayer of MoX$_2$, showing large oscillator strengths and spin-valley selection rules analogous to the monolayer excitons \cite{Urbaszek, Koch}. In addition, natural separation between the layers here is close to the threshold $d_c$, the latter being on the order of the exciton Bohr radius \cite{RP}. The electric-field tuning of the resonance position in this case of fixed $d$ may be possible due to suppression of the satellite carrier wavefunction in one of the layers (the residue of the intra-layer exciton) \cite{Urbaszek}.       

To conclude, we predict anomalously large enhancement of repulsive interactions in a system of dipolar polaritons. The proposed model is based on the physics of a bound state separated from the outer continuum by a potential barrier. Our results apply to a wide variety of 2D semiconductor heterostructures, such as atomically thin layers of TMD's and quantum wells. An intriguing prediction of our theory is that the resonantly paired polaritons represent an efficient source of squeezed radiation. This might be readily verified by examining the statistics of emitted photons with the balanced homodyne detection \cite{homodyne}. The idea of using the shape resonance to produce strong pair correlations and squeezing at ultra-low polariton densities opens wide perspectives for future research and applications. Thus, an interesting new direction would be application of the physics discussed in this work to the recently established field of topological polaritons \cite{Nalitov, topological}.                        

The author acknowledges the support by Russian Science Foundation (Grant No. 18-72-00013). I also thank Misha Glazov for helpful reading of the manuscript.

\section*{Supplemental material}

\subsection{Polariton Fano-Anderson model}

Consider 2D excitons coupled to photon modes in a planar microcavity of width $L_c$ and area $S$ with perfect mirrors. The photon dispersion reads
\begin{equation}
\hbar\omega_\mathrm{ph}(\bm p)=\hbar c\sqrt{p^2+\pi^2/L_c^2},
\end{equation}
that at low $\bm p$ has a massive form
\begin{equation}
\label{Eph}
\hbar\omega_\mathrm{ph}(\bm p)=E_\mathrm{ph}+\frac{\hbar^2 p^2}{2m_\mathrm{ph}}
\end{equation}
with $m_\mathrm{ph}=\pi^2\hbar/cL_c$. The exciton dispersion is given by
\begin{equation}
\label{Ex}
\hbar\omega_\mathrm{X}(\bm p)=E_\mathrm{X}+\frac{\hbar^2 p^2}{2m_\mathrm{X}},
\end{equation}
and in typical all-dielectric microcavities one has
\begin{equation}
\label{MassRatio1}
m_\mathrm{X}\gg m_\mathrm{ph}.
\end{equation} 
Thus, for GaAs-based structures $m_\mathrm{X}\sim 10^4 m_\mathrm{ph}$.

The prototype Hamiltonian may be recast as a sum of two contributions $\hat H=\hat H_\mathrm{sp}+\hat V$, where
\begin{equation}
\label{SP}
\begin{split}
\hat H_\mathrm{sp}&=\sum_{\textbf{p},\sigma}\hbar\omega_\mathrm{ph}(\bm p)\hat a_{\sigma, \textbf{p}}^{\dag} \hat a_{\sigma, \textbf{p}}+\sum_{\textbf{p},\sigma}\hbar\omega_\mathrm{X}(\bm p)\hat b_{\sigma, \textbf{p}}^{\dag} \hat b_{\sigma, \textbf{p}}\\
&+\sum_{\textbf{p},\sigma}i\hbar g(\bm p)(\hat b_{\sigma, \textbf{p}}^{\dag}\hat a_{\sigma, \textbf{p}}-\hat b_{\sigma, \textbf{p}}\hat a_{\sigma, \textbf{p}}^{\dag})
\end{split}
\end{equation}
is the single-particle term describing formation of exciton-polaritons (see below) and
\begin{equation}
\label{2body}
\hat V=\frac{1}{2S}\sum_{\textbf p_1,\textbf p_2,\textbf{q},\sigma,\sigma^\prime}\hat b_{\sigma, \textbf p_1+\textbf q}^{\dag} \hat b_{\sigma^\prime,\textbf p_2-\textbf q}^{\dag}V_{\sigma \sigma^\prime}(\bm q)\hat b_{\sigma, \textbf p_1}\hat b_{\sigma^\prime,\textbf p_2}
\end{equation}
being the two-body interaction between the excitons,
\begin{equation}
V_{\sigma\sigma'}(\bm q)=\int e^{-i\bm q\bm r}V_{\sigma\sigma'}(\bm r)d\bm r.
\end{equation}
We shall be particularly interested in $V_{\uparrow\downarrow}(\bm r)$, schematically shown by the solid black line in Fig. \ref{Fig1}. At large distances this potential is identical to $V_{\uparrow\uparrow}(\bm r)$ and $V_{\downarrow\downarrow}(\bm r)$, and is characterized by the dipolar tail $\hbar^2 r_\ast/m r^3$, where
\begin{equation}
r_\ast=\frac{m e^2 d^2}{\kappa \hbar^2}
\end{equation}
is the dipolar length. At $r\sim r_\ast$, however, it features a potential well. At sufficiently small $d$ this potential well admits a true bound state - biexciton. Evolution of this bound state into a resonance as one increases $d$ was described in \cite{RP}. Here we aim at extending that phenomenology to polaritons.

\begin{figure}[t]
\includegraphics[width=1\columnwidth]{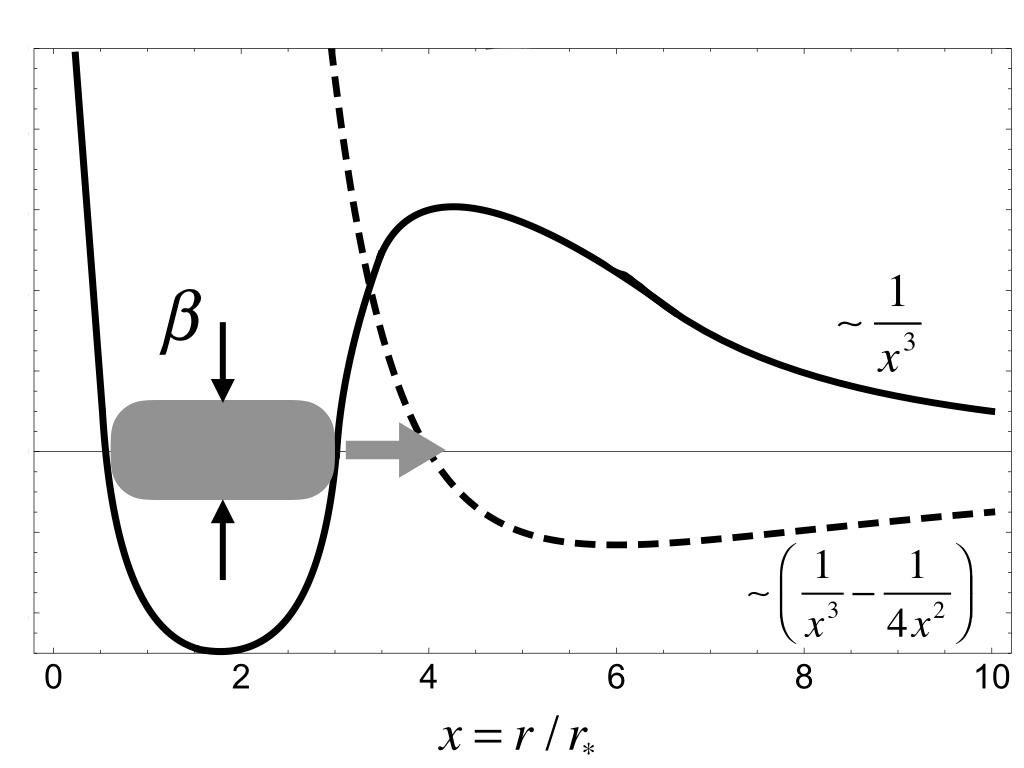}
\caption{Schematic illustration of the two-body exciton interaction potnetial $V_{\uparrow\downarrow}(r)$ (bold line). Due to the 2D kinematics the effective 1D potential felt by the radial wavefunction of the relative motion of excitons (dashed line) has an attractive tail which strongly suppresses the dipolar potential barrier. As a result, the tunneling amplitude is greatly enhanced. For polaritons this yields exponential suppression of the ratio of the resonance energy $\varepsilon$ to its width $\beta$, Eq. \eqref{WidthEstimate}.}
\label{Fig1}
\end{figure} 

The standard procedure is to diagonalize the single-particle term \eqref{SP} by doing the Hoppfield transformation
\begin{equation}
\hat{p}_{\sigma, \bm p}=u_{\bm p} \hat b_{\sigma,\bm p}+\upsilon_{\bm p} \hat a_{\sigma,\bm p}
\end{equation}
with $\lvert u_p\rvert^2+\lvert \upsilon_p\rvert^2 =1$ in order to preserve the boson commutation relation. One gets two eigenvalues $E_\pm(\bm p)$  corresponding to the upper and lower polariton branches. The Hoppfield coefficients are related by
\begin{equation}
u_{\sigma,\bm p}^{\pm}=\frac{i \hbar g\upsilon_{\sigma,\bm p}^{\pm}}{\hbar\omega_\mathrm{X}(\bm p)-E_\pm(\bm p)}.
\end{equation}
In the long-wavelength limit $\bm p\rightarrow 0$ we may take $u_{\bm p}^{\pm}\approx u_{0}^{\pm}$ and $\upsilon_{\bm p}^{\pm}\approx \upsilon_{0}^{\pm}$. Furthermore, assume $E_\mathrm{ph}=E_\mathrm{X}$ in Eqs. \eqref{Eph} and \eqref{Ex}. We get
\begin{equation}
\hat{p}_{\sigma,\bm p}^{\pm}=\frac{\hat b_{\sigma,\bm p}\pm\hat a_{\sigma,\bm p}}{\sqrt{2}}
\end{equation}
and
\begin{equation}
E_\pm(0)=E_\mathrm{X}\pm \hbar g.
\end{equation}
One may also estimate $m_{+}=m_{-}\equiv m\approx 2m_\mathrm{ph}$. By virtue of \eqref{MassRatio1}
\begin{equation}
\label{MassRatio2}
m\ll m_\mathrm{X}.
\end{equation}
We define
\begin{subequations}
\begin{align}
\hat c_{\sigma,\bm p}&\equiv \hat{p}_{\bm p}^{-}\\
\hat d_{\sigma,\bm p}&\equiv \hat{p}_{\bm p}^{+}
\end{align}
\end{subequations}
for the lower and upper polaritons, respectively. In terms of the new operators the single-particle Hamiltonian \eqref{SP} reads
\begin{equation}
\hat H_\mathrm{sp}=\sum_{\textbf{p},\sigma}E_{-}(\bm p)\hat c_{\sigma, \textbf{p}}^{\dag} \hat c_{\sigma, \textbf{p}}+\sum_{\textbf{p},\sigma}E_{+}(\bm p)\hat d_{\sigma, \textbf{p}}^{\dag} \hat d_{\sigma, \textbf{p}}.
\end{equation}
By substituting
\begin{equation}
\hat b_{\sigma,\bm p}=\frac{\hat c_{\sigma,\bm p}+\hat d_{\sigma,\bm p}}{\sqrt{2}}
\end{equation}
into Eq. \eqref{2body} and assuming occupation of the lower polariton branch only, one gets
\begin{equation}
\label{2body}
\hat V=\frac{1}{8S}\sum_{\textbf p_1,\textbf p_2,\textbf{q},\sigma,\sigma^\prime}\hat c_{\sigma, \textbf p_1+\textbf q}^{\dag} \hat c_{\sigma^\prime,\textbf p_2-\textbf q}^{\dag}V_{\sigma \sigma^\prime}(\bm q)\hat c_{\sigma, \textbf p_1}\hat c_{\sigma^\prime,\textbf p_2}.
\end{equation}
Thus, one can see that the low-momentum scattering of polaritons occurs in the same potential (rescaled by a numerical factor) as the scattering of excitons, but with a drastically reduced effective mass \eqref{MassRatio2}. By changing notations $ E_{-}(\bm p)\equiv E(\bm p)$ and repeating the arguments of Ref. \cite{RP}, one may derive the Fano-Anderson model (2) used in the main text. The predictions and physical meaning of this model depend on the ratio $\varepsilon/\beta$. Below we shall argue that the relation \eqref{MassRatio2} implies $\varepsilon/\beta\ll 1$, so that quite generally one may expect the polaritons to interact via a broad resonance.

\subsection{Dissociation of a dipolar bipolariton in two dimensions}

As we have established in the previous section, the polariton-polariton interaction potential apart of a numerical factor coincides with the bare exciton potential. For the $\uparrow\downarrow$ scattering channel of interest here this potential is schematically illustrated in Fig. \ref{Fig1}. We first note, that the nominal huge potential barrier becomes effectively suppressed due to the particularities of the 2D kinematics. To see this, consider the radial 2D Schrodinger equation for the relative motion with zero angular momentum ($s$-wave):
\begin{equation}
\label{sWave}
\left (\frac{\partial^2}{\partial r^2}+\frac{1}{r}\frac{\partial}{\partial r}-\frac{m}{\hbar^2}V_{\uparrow\downarrow}(r)+k^2\right)\phi(r)=0
\end{equation}
The dissociation rate of $\phi(r)$ may be estimated by considering the tunnelling in the effective one-dimensional model
\begin{equation}
\frac{\hbar^2}{m}\chi^{\prime\prime}+[E-V_\mathrm{eff}(r)]\chi=0,
\end{equation}
which is obtained from \eqref{sWave} by substituting $\phi(r)=\chi(r)/\sqrt{r}$. Here
\begin{equation}
V_\mathrm{eff}(r)=V_{\uparrow\downarrow}(r)-\frac{\hbar^2}{4m r^2}\approx \frac{\hbar^2}{mr_\ast^2}\left(\frac{1}{x^3}-\frac{1}{4x^2}\right),
\end{equation}
where we have taken $V_{\uparrow\downarrow}(r)\approx \hbar^2 r_\ast/m r^3$ and defined $x=r/r_\ast$. This potential is shown by the dashed line in Fig. \ref{Fig1}. One can see that the tunnelling at $E=0$ occurs in a narrow region from $r_a\approx r_\ast$ to $r_b=4r_\ast$. For a crude estimate we may use the quasiclassical approximation \cite{Landau} to calculate the tunnelling probability as a product of the number of the particle beatings per unit of time $\sim p/mr_\ast\sim\hbar/mr_\ast^2$ and the barrier transmission
\begin{equation}
D(E)=\exp\left(-\frac{2}{\hbar}\int\limits_{r_a}^{r_b}\sqrt{m(V_\mathrm{eff}(r)-E)}dr\right).
\end{equation}   
By evaluating the integral one gets $D(0)\sim 0.25$. Thus, we may write
\begin{equation}
\beta\sim\frac{\hbar^2}{m r_\ast^2}.
\end{equation}
Hence, we see that, by virtue of \eqref{MassRatio2}, the bipolariton resonance width should greatly exceed the analogous quantity for bare excitons $\beta_{\mathrm X\mathrm X}$,
\begin{equation}
\frac{\beta}{\beta_{\mathrm X\mathrm X}}\sim\frac{m_{\mathrm X}}{m}\gg 1.
\end{equation}
On the other hand, by using the formula
\begin{equation}
\varepsilon=-\frac{\hbar^2}{ma^2},
\end{equation}
where
\begin{equation}
a\sim r_\ast e^{\Delta d/(d_c-d)}
\end{equation}
and taking into account that $\Delta d\sim \beta$ \cite{RP}, we may write
\begin{equation}
\frac{\varepsilon}{\varepsilon_{\mathrm X\mathrm X}}\sim\frac{m_{\mathrm X}}{m}\exp\left(-\frac{m_{\mathrm X}}{m}\frac{\Delta d_{\mathrm X\mathrm X}}{d_c-d}\right).
\end{equation}            
Thus, even if one has $\Delta d_{\mathrm X\mathrm X}\sim (d_c-d)$ and, consequently,  $\varepsilon_{\mathrm X\mathrm X}\sim \beta_{\mathrm X\mathrm X}$ for the bare excitons, for the polaritons one would obtain
\begin{equation}
\label{WidthEstimate}
\frac{\varepsilon}{\beta}\sim e^{-m_{\mathrm X}/m}\ll 1.
\end{equation}
We conclude that a dipolar "bipolariton" may exist only as a \textit{broad resonance}. This result generalizes the earlier estimate of the bipolariton binding energy at $d=0$ (non-dipolar excitons) \cite{Rocca}.

\end{document}